\documentclass[journal,table,xcdraw]{IEEEtran}
\usepackage{cite}
\usepackage{amsmath,amssymb,amsfonts}
\usepackage{graphicx}
\usepackage{textcomp}
\usepackage{xcolor}
\usepackage{cite}
\usepackage{hyperref}
\usepackage{multirow}
\usepackage{arydshln}
\usepackage{graphicx}
\usepackage{svg}
\usepackage{subcaption}
\usepackage{cleveref}
\usepackage{algorithm}
\usepackage{algpseudocode}
\usepackage{gensymb}
\usepackage{amsmath}
\usepackage{nomencl}
\makenomenclature
\usepackage{booktabs,tabulary}
\usepackage{xurl}

\hypersetup{ 
	colorlinks=true,
	linkcolor=black,
	filecolor=black,      
	urlcolor=black,
	citecolor=black,
}

\def\BibTeX{{\rm B\kern-.05em{\sc i\kern-.025em b}\kern-.08em
    T\kern-.1667em\lower.7ex\hbox{E}\kern-.125emX}}

\begin{document}

\title{A Preventive-Corrective Scheme for Ensuring Power System Security During Active Wildfire Risks}



\author{\IEEEauthorblockN{Satyaprajna Sahoo, \textit{Student Member, IEEE}, and
Anamitra Pal, \textit{Senior Member, IEEE}}

\thanks{This work was supported in part by  the National Science Foundation (NSF) grant under Award ECCS-2132904.

Satya Sahoo (email: sssahoo2@asu.edu) is a graduate student, and Anamitra Pal (email: Anamitra.Pal@asu.edu) is an Associate Professor in the School of Electrical, Computer, and Energy Engineering (ECEE) at Arizona State University (ASU).
}

\vspace{-2.5em}
}


\maketitle

\begin{abstract}
The focus of this paper is on operating the electric power grid in a secure manner {during scenarios of active wildfire risk}.
This is a challenging problem
because of the uncertain ways in which the fires 
can impact the operation of the power system.
To address this challenge, we propose
a novel preventive-corrective coordinated
decision-making scheme
that quickly mitigates both static and dynamic insecurities 
given the risk of active wildfires in a region. 
The scheme utilizes a
comprehensive contingency analysis tool 
for \textit{multi-asset outages} that leverages:
(i) a
``Feasibility Test" algorithm which exhaustively desaturates overloaded cut-sets to prevent cascading line outages, and (ii) a data-driven transient stability analyzer which alleviates dynamic instabilities.
This tool
is then used to operate a coordinated unit commitment/optimal power flow model that is designed to adapt to varying risk levels associated with wildfires. Depending on the allowed
risk, the model 
balances economical operation and grid robustness. 
The results obtained using the IEEE 118-bus system  
indicate that the proposed approach alleviates 
system vulnerabilities to wildfires while also minimizing 
operational cost.
\end{abstract}

\begin{IEEEkeywords}
Cut-set saturation, Preventive-corrective coordination, Static security, Transient stability, Wildfire
\end{IEEEkeywords}

\IEEEpeerreviewmaketitle
\printnomenclature[1.3cm]

\section{Introduction}


\IEEEPARstart{I}{n} recent times, the increasing frequency, intensity, and spread of wildfires has presented significant challenges not only to emergency personnel and forest authorities, but also to the operation of electric power systems.
While the impacts on emergency services and forest officials are relatively easier to understand, the interaction between wildfires and the electric power infrastructure is more complex.
Power utilities are expected to take measures to prevent initiation of wildfires by 
their equipment (most often, power lines),
while simultaneously operating the grid in a robust and economic manner
in the presence of high wildfire risks \cite{9677975,muhs2020wildfire,nazemi2022powering,umunnakwe2023data}.
While past research has extensively explored the former issue, this paper focuses on the latter: \textit{ensuring security and stability of the power system in the face of active wildfire risks}.

Analyses of major 
power system interruptions related to wildfires have identified cascading outages resulting from overloaded lines, 
instabilities caused by
frequent arc-faults, and/or preemptive disconnections of the lines
as the primary cause(s) (for the interruptions).
Overloads occur because the heat from ongoing wildfires affect the nearby power lines
resulting in \textit{lowering of the conductor's current carrying capacity} \cite{rostamzadeh2023optimal}. This lowering can then cause bottlenecks to appear in other parts of the system 
\cite{CHOOBINEH201520}.
%
Arc-faults are not only precursors of grid-initiated wildfires but also they occur when a fire ignited by another source (e.g., lightning) approaches the power lines.
Wildfire inducing/induced arc-faults are unique in the sense that
they
occur \textit{multiple times within a 
short time-period
} \cite{north20161}. Conventional
contingency analysis tools that usually deal with 
non-repeating faults
are not equipped to handle such phenomena.
It is crucial that both the static security as well as the dynamic stability implications of a potential wildfire contingency are considered for ensuring reliable and resilient power system operation.

{To prevent grid-induced fires and protect grid assets from wildfires, grid operators often resort to public safety power shutoffs (PSPS), in which power lines are preemptively disconnected. Typically, PSPS is carried out heuristically based on operators' experience \cite{PSPS_Workstream_2021}.
However, from a practical perspective, 
\textit{preemptively disconnecting lines bears a very high social cost} \cite{9737413}. 
For example, over 2.7 million people lost electricity in California in 2019
when preemptive measures taken by the local utility resulted in more than 940,000 homes and businesses losing electricity \cite{Columbia}. While PSPS is effective at minimizing grid induced fires, suboptimal implementation may lead to significant power outages during wildfire scenarios, as network de-energization affects power flows in the remainder of the network, leading to \textit{potential cascading outages}.}

While there is significant research aimed at wildfire risk management in the context of the electric power grid
\cite{9409078},
these efforts predominantly revolve around 
risk quantification \cite{9840510},
resilience assessment \cite{donaldson2022integration}
resource allocation \cite{10164135}, and/or 
system upgrades \cite{10032578}. 
Addressing the security and stability challenges posed by wildfires in real-time requires a holistic
approach that is currently missing in the state-of-the-art (\cite{rostamzadeh2023optimal,CHOOBINEH201520,north20161,9737413,9409078,9840510,donaldson2022integration,10164135,10032578}). 
{Mitigating wildfire risk-induced security concerns \textit{after} the wildfire had started was explored in  \cite{YANG2024110773,salimi2024equitable,nazemi2022resilient,yang2024multi}, which treated it as an operations research problem.}
At the same time, static security assessment and enhancement in the context of storms and hurricanes with the aim of minimizing operational cost and load-shed has been formulated in \cite{biswas2020graph,abdelmalak2022proactive,9420300}.
Similarly, a  multi-state model for dynamic security enhancement against faults caused by hurricanes was developed in \cite{9290096}.
{However, \cite{YANG2024110773,salimi2024equitable,nazemi2022resilient,yang2024multi,biswas2020graph,abdelmalak2022proactive,9420300,9290096}
either assessed stability after contingency manifestation or did not utilize
the stability results in both the day-ahead and the real-time stages.} 
Specifically, coordination between day-ahead and real-time operations is crucial for achieving robust \textit{and} economic operation when multiple contingencies manifest.

Lastly, although prior research has integrated transient stability constraints into an optimal power flow (OPF) problem (called TSCOPF henceforth) \cite{yuan2020preventive,sevilla2023two}, or a unit commitment (UC) problem (called TSCUC henceforth) \cite{wu2023transient}, such problems have not been investigated in the presence of active wildfire risks. 
Moreover, the \textit{direct} application of TSCOPF/TSCUC to an extreme event scenario (such as wildfire) is not appropriate because of the heightened uncertainty associated with such scenarios, as well as the severely limiting nature of the stability constraints. 
{Our prior work \cite{sahoo2023cut} introduced a cut-set and stability constrained OPF algorithm which assumed 
that the wildfire had already started.
However, scenarios of active wildfire risks, which are more prevalent, was not analyzed in that study. Furthermore, preparation for periods of high risk in the day-ahead stage and coordination with varying risk scenarios in real-time, was also not considered in \cite{sahoo2023cut}. 
This paper addresses the above-mentioned limitations by making the following salient contributions: 
}

\begin{figure*}[ht]
\centering
\includegraphics[width=1\textwidth]{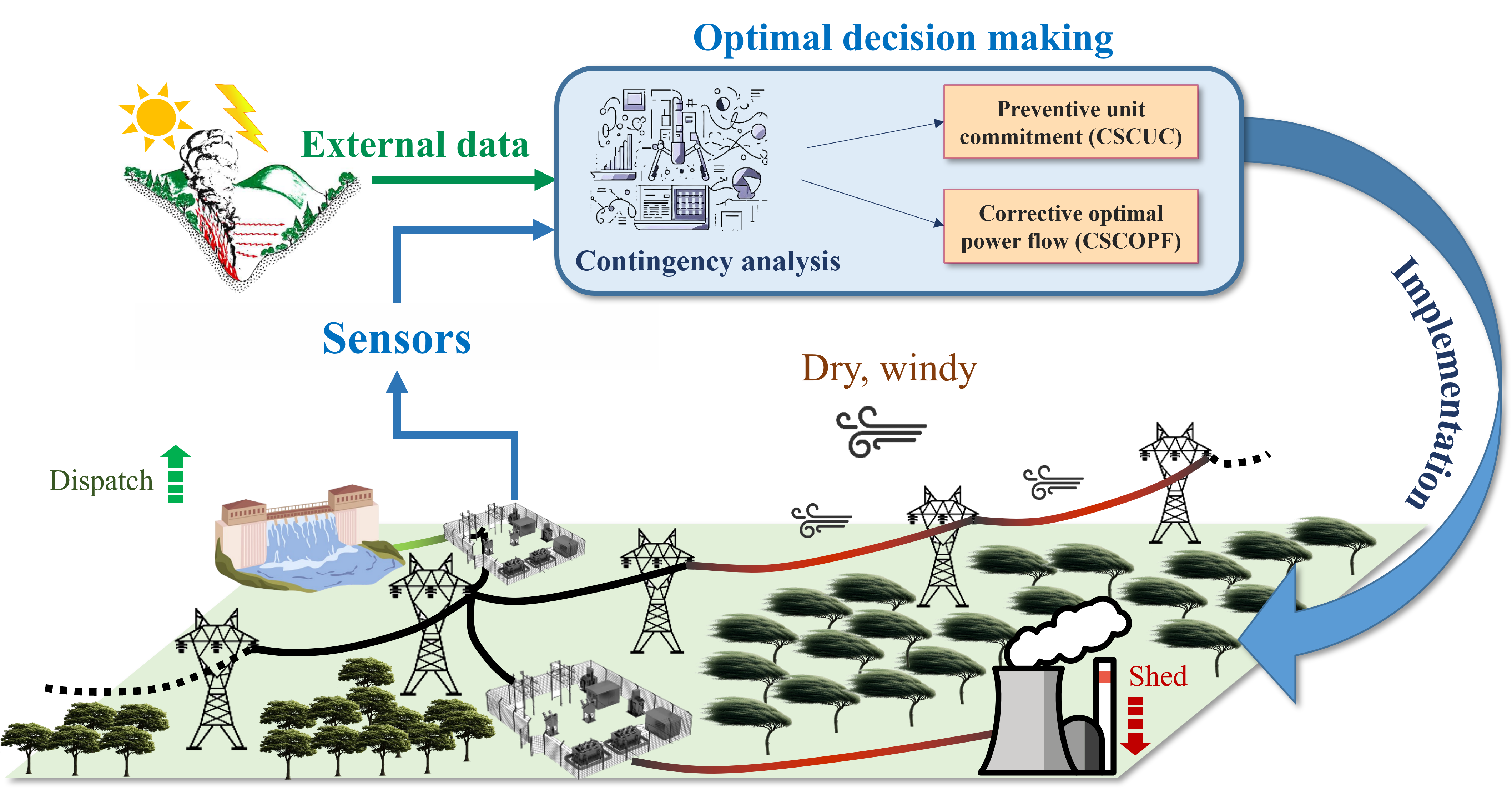}
\caption{\textcolor{black}{Overview of the proposed preventive-corrective coordinated action scheme.}}
\label{fig:PCoverview}
\end{figure*}
\begin{itemize}
    \item {A comprehensive contingency analysis tool for wildfire inducing/induced faults initially introduced in \cite{sahoo2023cut} is utilized to exhaustively analyze \textit{multi-asset outages}, considering the ability of these outages to affect the grid from both static security as well as dynamic stability perspectives.
    The tool first leverages
    a \textit{Feasibility Test} (FT) \cite{9420300} algorithm that, given a sequence of outages, is able to quickly and exhaustively identify and correct cut-set\footnote{A cut-set is a set of lines, which if tripped, would create disjoint islands in the network. Therefore, saturated/overloaded cut-sets are the most vulnerable interconnections of the system as they have limited power transfer capability. For more details, refer to \cite{9420300}.} saturation.
    Secondly, the tool
    has a transient stability analysis 
    component
    that utilizes machine learning (ML) to analyze and correct rotor angle instabilities caused by  multiple frequent arc-faults and line outages.}

    \item {This contingency analysis tool
    is then used to operate a 
    two-stage 
    preventive-corrective coordinated optimization model to optimally allocate resources based on actual system conditions and allowed risk.
    The preventive component is based on a cut-set and stability constrained UC (CSCUC) that brings additional generators online to support those regions of the system that are at risk of wildfires in the near-future.}
    
    \item {The corrective component is based on a 
    cut-set and stability constrained OPF (CSCOPF) which is extended from our previous work in \cite{sahoo2023cut}
    by taking real-time wildfire risk and load forecasts into account to provide a wider range of solutions that balance between system robustness and operational cost. 
    }

    
\end{itemize}

{
The proposed model is implemented on simulated wildfire scenarios for the IEEE 118-bus system. A comparison with the conventional real-time security constrained economic dispatch (RT-SCED), a traditional TSCOPF model \cite{10138159}, and our previous work on CSCOPF \cite{sahoo2023cut}
indicates that the approach developed in this paper gives demonstrably better results.
A sensitivity analysis is conducted next to gain additional operational insights. Finally, a case-study is performed to emphasize the contributions of the proposed model in severe islanding cases 
including its 
ability to facilitate PSPS.}
The rest of the paper is structured as follows. 
Section \ref{Section2} first discusses the impact of wildfires on power system operations, and then proposes two solution methodologies for mitigating the steady-state and transient impacts of such extreme events.
Section \ref{Section3} integrates these two methodologies into a UC/OPF formulation to create a coordinated 
decision-making framework that considers the risk of a wildfire event in the region for optimally operating the system.
{
Section \ref{Section4} entails the implementation of the proposed model on  the IEEE 118-bus system by simulating a wildfire scenario on an identified vulnerable corridor of that system. The section also considers extreme scenarios such as islanding as well as describes the integration of the proposed model with PSPS operations.}
The conclusions are drawn in Section \ref{Section5}.

\section{Strategies to Mitigate Impacts of Multi-asset Outages During Wildfires}
\label{Section2}

\subsection{Problem Scope}
\label{sec:1a}

Wildfires and their interaction with the electric power infrastructure constitute a multi-faceted problem. 
Grid-initiated fires are usually the result of high-impedance faults that: (a) occur under environmental conditions that are conducive for a faster spread, (b) are harder to detect, and (c) happen in quick succession \cite{9305959,9449670,10042433,
8253878}.
The fear is so severe that utilities 
are often forced to change 
their protection settings apriori (say, a day before) 
in anticipation of high wildfire risks.
For fires started by another source and approaching the power lines: 
(a) the spatio-temporal process of wildfire spread is based on the local climate and geography (topography, vegetation, wind); 
(b) the breakdown mechanisms of the air gap around the lines vary with time, location, and wildfire proximity and intensity; 
(c) the outcomes can range from multiple arcing events to a complete line melt-down (permanent outage)
\cite{ma2020real,daochun2015review,8620983}.
A variety of tools already exist for tracking wildfire spread over different geographical regions 
(e.g., FlamMap \cite{flammap}).
{This study assumes that the stochasticity of wildfire eruption, and the spread of that fire across neighboring regions can be expressed through a quantified risk metric for a given region (based on existing tools), and is available apriori (e.g, a day in advance).}

{From a power systems operations perspective, an economically optimal way to perform PSPS based on the knowledge of wildfire risk, termed optimal power shutoffs, has been introduced recently \cite{9305959}. 
However, it did not consider the dynamic impact of wildfires since these impacts involve assets that are usually not under the purview of PSPS. 
As such, a purely economics-focused scheme may end up worsening grid vulnerability from a stability view-point by increasing
loading on an already vulnerable area of the network.
Similarly, a 
scenario where a non-grid induced wildfire manifests unnoticed, may lead to the same outcome.} 
Therefore, this study is aimed at striking a balance between 
economical operation and grid 
stability
by pre-allocating resources in the day-ahead stage and reallocating them, if need be, in real-time. 

We now outline two power system vignettes that exemplify the scope of the methodology developed in this paper:

\noindent \textit{Vignette 1:} The risk of wildfire is high for the next day in a particular region but the fire has not started yet, and the power utility wants to pre-allocate/reallocate resources so as to secure the system without preemptively deenergizing all the power lines in that region.

\noindent \textit{Vignette 2:} A fire is burning in a neighboring region, and there is {a reasonably high}
probability of the fire spreading to the region-under-study on the following day. In this case, the utility wants to pre-allocate/reallocate
resources so that the system runs in a secure and stable manner while considering the dynamic nature of wildfire risks. 

An illustration
of the implementation of the proposed preventive-corrective coordinated action scheme is given in Fig. \ref{fig:PCoverview}. 
In the pre-contingency stage, a contingency analysis is performed on the high-risk areas, following which power flowing through that region is lowered and
additional generators are brought online in other regions (indicated by red and green arrows).
In the real-time stage, the online generators are optimally redispatched to quickly and economically steer the system to a secure and stable state.

Lastly, note that the proposed formulation is generic and, with suitable modifications, can be used to tackle other extreme events/multi-asset outage causing contingencies. The focus of this paper, however, is on wildfires.
In the following sub-section, we highlight the FT algorithm and the ML-based transient stability analysis which form the backbone of the proposed comprehensive contingency analysis tool for combating active wildfire risks.


\subsection{Feasibility Test (FT)-based Cut-set Security Analysis}
Since wildfires generally impact more than one line
in a target area,
their outage
can lead to a load-generation mismatch in areas that 
are connected by those lines.
Cut-sets, by definition, are the set of lines that join two areas. Therefore, by detecting and alleviating saturated cut-sets, security criteria during multiple line losses can be strengthened.
A saturated cut-set is one whose aggregate power flow exceeds the limits of the lines that form the cut-set. This is mathematically described by:
\begin{equation}
    \sum_{\forall e \in K_{\mathrm{crit}}} f_e > \sum_{\forall e \in K_{\mathrm{crit}}} f^{\mathrm{max}}_e
\end{equation}
where, $f^{\mathrm{max}}_e$ is the maximum power flow allowed through the $e^{th}$ line of the saturated cut-set, $K_{\mathrm{crit}}$.
By exploiting the principle that the cut-set power flow is not impacted by the method employed to redirect the
power flowing through a line that faces an outage \cite{biswas2020graph}, a fast and scalable algorithm called FT was developed that
exhaustively identifies and alleviates all saturated cut-sets for a given contingency \cite{9420300}.
The identified cut-sets can be desaturated by reducing the aggregate power flowing through the constituent lines. This is mathematically expressed as:
\begin{equation}
\label{eq:model_cut_set_constraint}
    \sum_{\forall e \in K_{\mathrm{crit}}} \Delta f_e \leq - \Delta P_{K_{\mathrm{crit}}} \quad  \forall K_{\mathrm{crit}} \in \kappa_{\mathrm{crit}}
\end{equation}
where, $\Delta P_{K_{\mathrm{crit}}}$ is the required transfer margin for $K_{\mathrm{crit}}$, and indicates the total amount of power that must be reduced across the lines of $K_{\mathrm{crit}}$ to prevent it from being overloaded.
{More details about the FT algorithm and its capabilities can be found in \cite{biswas2020graph,9420300}.}

\subsection{Machine Learning (ML)-based Transient Stability Analysis}
\label{ML-TSA}
Multiple arc-faults during periods of high wildfire risks and cascading outages 
caused by ongoing wildfires possess the ability to cause transient instability in the form of rotor angle instabilities in the synchronous generators present in the system. 
Transient stability (rotor angle stability) is assessed from the maximum difference in the rotor angles ($\delta_{\mathrm{max}}$) between two consecutive synchronous machines.
The metric used to analyze transient stability under such situations, called the transient stability index ($\mathrm{TSI}$), is mathematically defined by:
\begin{equation}
    \label{eq:eeac1}
    \mathrm{TSI} = \frac{360 - \delta_{\mathrm{max}}}{360 + \delta_{\mathrm{max}}} \times 100
\end{equation}

\nomenclature{\(\mathrm{TSI}\)}{Transient Stability Index}
\nomenclature{\(\delta\)}{Rotor angle}

The system is stable if $\mathrm{TSI}>0$, and unstable otherwise.
An unstable system is characterized by at least one generator losing sychronism, and eventually tripping.
For an unstable contingency, the total generation can be classified into stable and unstable generators, with the stable generators being those whose rotor angles are below $\delta_{\mathrm{max}}$.
Given a contingency that leads to transient instability, 
the instability can be corrected using the integrated extended equal area criterion (IEEAC) \cite{7395386}.
The theory of the IEEAC stipulates that transferring requisite amount of generation from the unstable generators ($\mathrm{CM}$) to the stable generators ($\mathrm{NM}$) can make the $\mathrm{TSI}$ positive. The required amount of generation that must be shifted is given by \cite{sahoo2023cut}:
\begin{equation} 
    \label{equation:Pshift}
    \Delta P_{tr} \geq \left(\frac{-\eta_{us} + \epsilon}{\tau_n} \right).\left(\frac{M}{M_{\mathrm{CM}}} + \frac{M}{M_{\mathrm{NM}}}\right)^{-1}
\end{equation} 
where, $M$, $M_{\mathrm{CM}}$, and $M_{\mathrm{NM}}$ are the one machine infinite bus (OMIB) inertia coefficients of the whole system, $\mathrm{CM}$, and $\mathrm{NM}$, respectively, and $\mathrm{CM} \cup \mathrm{NM} = G$.
$\Delta P_{tr}$ is referred to as the transient stability correction factor (TSCF).
Eq. \eqref{equation:Pshift} can be used to provide a transient stability constraint as shown below:
\begin{equation}
\label{eq:trans_stab_const}
    \sum_{\forall i \in \mathrm{CM}} \Delta p_{i} \leq - \Delta P_{tr}
\end{equation}

\nomenclature{\(M\)}{Inertia coefficient}
\nomenclature{\(\mathrm{NM}\)}{Set of non-critical machines}
\nomenclature{\(\eta_{us}\)}{Stability margin of the unstable system}
\nomenclature{\(\epsilon\)}{Desired stability margin}
\nomenclature{\(\tau_n\)}{Linear sensitivity factor}

Now, the calculation of $\Delta P_{tr}$ requires performing multiple time domain simulations (TDSs) for a single contingency, which is computationally expensive to do in real-time when
the list of potential contingencies is large.
Furthermore, one must also consider the variability of the loads - a factor that will be compounded if solar or wind is present. 
To account for both of these factors, we exploit the quasi-linear relationship between the pre-contingency one machine mechanical power and the transient stability margin \cite{7395386}.
Specifically, \textit{we posit that a quasi-linear relationship exists between the TSCF
and the pre-contingency loading condition ($l$)}, since the pre-contingency mechanical power is linearly related to the loads.
Consequently, we formulate a
linear regression
model that estimates the required TSCF for a forecasted loading condition, as shown below \cite{sahoo2023cut}:
\begin{equation}
    \Delta \hat{P}_{tr} = \sum_{i \in L} \theta_i l_i + \theta_0  = \Upsilon(l_i)
 \end{equation}
 \begin{equation}
     J(\theta) = \frac{1}{k} \sum_{i=1}^k (\Delta \hat{P}_{tr_i} - \Delta {P}_{tr_i})^2
 \end{equation}
where, $\theta$ are the weights, $J(\theta)$ is the loss function, and $k$ is the batch size.
This data-driven model for estimating the TSCF is referred to as the transient stability constraint prediction (TSCP) algorithm, and is denoted by $\Upsilon$.
The ability of different ML techniques in estimating $\Delta P_{tr}$ is compared in Section \ref{Section4A}.
{More details about the TSCP algorithm and its implementation can be found in \cite{sahoo2023cut}.}



With the FT algorithm and the ML-based transient stability analysis
generating appropriate
constraints, the goal is now to integrate them into the UC/OPF problem 
to achieve secure and stable power system operation during active wildfire risks, while also being 
economical. This goal is achieved through the preventive-corrective coordinated action scheme described in the next section.

\section{Preventive-corrective Coordination} \label{Section3}


In the day-ahead stage, the conventional UC formulation helps to find a low-cost operating schedule for the generators.
However, it can be suitably \textit{modified} to better prepare the system for extreme events. 
Similarly, the corrective actions of a \textit{modified} OPF formulation can minimize (if not eliminate) the impact of
contingencies by quickly reducing the power flowing through the more vulnerable areas in real-time.
Together, this coordinated preventive-corrective 
decision-making scheme can rapidly, safely, and economically steer the power system towards a secure and stable state.
The following sub-sections describe how such a scheme can be developed and implemented for managing active wildfire risks.

\vspace{-1em}
\subsection{Preventive Unit Commitment (UC)}
\label{sec:2b}

In the preventive stage, 
previously inactive generators are brought online to prepare the system for 
a contingency that might occur in the near-future. Note that although the generators are brought online, they may or may-not dispatch in real-time (this decision is made in the next sub-section).
Hence, we formulate a \textit{modified} UC problem as described below.
We start by expressing the generator costs $(F_i)$ as a quadratic function of the output power ($p_i$):
\begin{equation}
    F_i (p_{i}) =  {a_i + b_i  p_{i} +  c_i (p_{i})^2 }
\end{equation}

\nomenclature{\(F_i\)}{Cost function of generator $i$}
\nomenclature{\(p_i\)}{Active power output of generator $i$}
\nomenclature{\(a_i\)}{Static cost coefficient of generator $i$}
\nomenclature{\(b_i\)}{Linear cost coefficient of generator $i$}
\nomenclature{\(c_i\)}{Quadratic cost coefficient of generator $i$}

Since the proposed formulation is modeled as an optimal rescheduling problem, the cost of change in generation can be simplified to:
\begin{equation}
\label{eq6}
\begin{aligned}
\Delta F_i (\Delta p_{i}) = \quad &  \{a_i + b_i  p^n_{i} +  c_i (p^n_{i})^2 \}\\
 \quad &  - \{a_i + b_i  p^0_{i} +  c_i (p^0_{i})^2 \} \\
 \quad &  = c_i (\Delta p_{i})^2 + ( b_i + 2 c_i p^0_{i})\Delta p_{i}\\
\end{aligned}
\end{equation}
where, the superscripts $^0$ and $^n$ refer to the pre-contingency and post-contingency status, respectively, of the corresponding variable, and $\Delta p_{i} = p^n_{i} - p^0_{i}$. 
\nomenclature{\(p^0_i\)}{Pre-contingency active power produced by generator $i$}
\nomenclature{\(p^n_i\)}{Post-contingency active power produced by generator $i$}
The overall objective should include the cost of generation change and load shed while also considering the availability of inactive generators. 
Now, the UC decisions are based on the post-contingency operation (which generators are most economical after the vulnerable assets are out), hence
only the generators that are currently inactive are
considered in the modified UC.
This decreases the number of binary variables in the proposed formulation \textit{by a considerable amount}. 
Taking all of this into account, the desired objective function can be written as:
\begin{equation}
\label{eq:new_obj}
    \begin{aligned}
    \min_{\Delta p_{i}, \Delta l_j, u_i} \quad &  \sum_{\forall i \in G} (c_i \Delta p_{i}^2 + d_i \Delta p_{i}) + \sum_{\forall j \in L} (m_j \Delta l_j)\\
     \quad & + \sum_{\forall i \in G_d} u_i a_i\\ 
    \end{aligned}
\end{equation}
where, $d_i = (b_i + 2 c_i p^0_{i})$.
The set of inactive generators $G_d$ is a subset of 
$G$.
Load shed cost $m_j$ is chosen to be higher than the generator costs, to de-incentivize power outage. The binary variable $u_i$ is used to bring additional generators online 
considering the post-contingency situation, including the assets out of service due to the contingency.
The overall objective is now subjected to the following constraints:
\begin{equation}
\label{eq:lim_1}
    u_i . (p_{i}^{\mathrm{min}} - p_{i}^0) \geq \Delta p_{i} \geq u_i . (p_{i}^{\mathrm{max}} - p_{i}^0) \;  \forall i \in G
\end{equation}
\begin{equation}
\label{eq:lim_1.5}
    u_i = 1 \quad \quad \quad \forall i \in \{G - G_d\}
\end{equation}
\begin{equation}
    \label{eq:lim_2}
    l_j^{\mathrm{min}} - l_j^0 \geq \Delta l_j \geq l_j^{\mathrm{max}} - l_j^0 \quad \forall j \in L
\end{equation}
\begin{equation}
    \label{eq:lim_3}
\begin{aligned}
     \quad & f_e^{\mathrm{min}} - f_e^0 \leq \sum_{\forall i \in G} \mathrm{PTDF}^r_{e,i} \Delta p_{i} - \sum_{\forall j \in L} \mathrm{PTDF}^r_{e,j} \Delta l_j \\
      \quad & \leq f_e^{\mathrm{max}} - f_e^0   \quad \forall e \in B_r\\
\end{aligned}
\end{equation}
\begin{equation}
    \label{eq:lim_5}
    \sum_{\forall i \in G} \Delta p_{i} = \sum_{\forall j \in L} \Delta l_j
\end{equation}
\begin{equation}
\label{eq:lim_6}
\begin{aligned}
    \quad & \sum_{\forall i \in G} (\mathrm{PTDF}^r_{e,i} + \mathrm{LODF}_{e,k} \mathrm{PTDF}^r_{k,i}) \Delta p_{i}\\
     - \quad & \sum_{\forall j \in L} (\mathrm{PTDF}^r_{e,j} + \mathrm{LODF}_{e,k} \mathrm{PTDF}^r_{k,j}) \Delta l_j\\
    \leq \quad & f_e^{\mathrm{max}} - f_e^0 + (\mathrm{LODF}_{e,k} f_k^0) \quad \forall e,k \in B_r, \xi
\end{aligned}
\end{equation}
\begin{equation}
\label{eq:lim_7}
\begin{aligned}
    \quad & \sum_{\forall i \in G} (\mathrm{PTDF}^r_{e,i} + \mathrm{LODF}_{e,k} \mathrm{PTDF}^r_{k,i}) \Delta p_{i} \\
     - \quad & \sum_{\forall j \in L} (\mathrm{PTDF}^r_{e,j} + \mathrm{LODF}_{e,k} \mathrm{PTDF}^r_{k,j}) \Delta l_j\\
    \geq \quad & f_e^{\mathrm{min}} - f_e^0 + (\mathrm{LODF}_{e,k} f_k^0) \quad \forall e,k \in B_r, \xi
\end{aligned}
\end{equation}
\begin{equation}
\label{eq:lim_8}
    \begin{aligned}
        \quad & \sum_{\forall i \in G} (\sum_{\forall u \in K_{\mathrm{crit}}} \mathrm{PTDF}_{u,i}) \Delta p_{i} \\
        - \quad & \sum_{\forall j \in L} (\sum_{\forall u \in K_{\mathrm{crit}}} \mathrm{PTDF}_{u,j}) \Delta l_j\\
        \leq \quad &  -\Delta P_{K_{\mathrm{crit}}} \quad \forall K_{\mathrm{crit}} \in \kappa_{\mathrm{crit}}
    \end{aligned}
\end{equation}
\begin{equation}
\label{eq:trans_stab_const_lim_9}
    \sum_{\forall i \in \mathrm{CM}} \Delta p_{i} \leq - \Delta P_{tr}
\end{equation}
\nomenclature{\(f_e\)}{Active power flowing in branch $e$}
\nomenclature{\(B_r\)}{Set of all branches}
\nomenclature{\(\mathrm{CM}\)}{Set of critical machines}
\nomenclature{\( \Delta P_{tr} \)}{Transient stability transfer margin}
\nomenclature{\(G\)}{Set of all generators}
\nomenclature{\(L\)}{Set of all loads}
\nomenclature{\(m_j\)}{Cost of load-shed}
\nomenclature{\(G_d\)}{Set of inactive generators}
\nomenclature{\(u_i\)}{Binary variable signifying status of generator $i$}
\nomenclature{\(l_j\)}{Active power of load $j$}
\nomenclature{\(\xi\)}{Contingency List}
\nomenclature{\(\mathrm{PTDF}\)}{Power Transfer Distribution Factor}
\nomenclature{\(\mathrm{LODF}\)}{Line Outage Distribution Factor}
\nomenclature{\(\Delta P_{K_{\mathrm{crit}}}\)}{Transfer margin identified for cut-set $K_{\mathrm{crit}}$}
\nomenclature{\(K_{\mathrm{crit}}\)}{Identified saturated cut-set}
\nomenclature{\(\kappa_{\mathrm{crit}}\)}{Set of all identified saturated cut-sets}

Eqs. \eqref{eq:lim_1}-\eqref{eq:lim_3} provide limits for the optimization variables. Note that the binary variable $u_i$ is set to 1 for the already dispatched generators; this is indicated by \eqref{eq:lim_1.5}.
The power balance constraint
is shown in \eqref{eq:lim_5}. The post-contingency $\mathrm{N-1}$ branch overload constraints
are expressed in \eqref{eq:lim_6}-\eqref{eq:lim_7}. The post-contingency cut-set constraint is modeled in \eqref{eq:lim_8}, while
\eqref{eq:trans_stab_const_lim_9} denotes the transient stability constraint. 
This completes the description of the modified UC, which is henceforth referred to as the CSCUC.

Algorithm \ref{alg:CSCUC}  describes the implementation of CSCUC.
In the day-ahead stage, a list of contingencies are obtained, and contingency analysis is done with each list using TDS. If a contingency is found to cause transient instabilities or cut-set saturation, then the required transfer margins are calculated, and the TSCP is trained. This is followed by the CSCUC, which recommends additional generators to be brought online 
in anticipation of the contingency manifesting the next day.

\begin{algorithm}
\caption{CSCUC Implementation}\label{alg:CSCUC}
\textbf{Input:} Historical load distributions, Contingency list $\xi$ \\
\textbf{Output:} Unit Commitment  (UC) status $u_i$\\
\textbf{Day-ahead:}
\begin{algorithmic}[1]

\State Perform random sampling from the 
historical load distributions to determine potential loading conditions $\Xi$
\State Define empty list of constraints $\Phi$
\For{loading condition
in $\Xi$}
    \State Perform TDS using $\xi$
    \If{violations detected}
        \State Generate transient stability constraint using \eqref{equation:Pshift}
        \State Update $\Phi$
    \EndIf
\EndFor
\State Train $\Upsilon$ using $\Phi$

\State Run FT to generate constraint \eqref{eq:lim_8}


\State CSCUC: Set objective \eqref{eq:new_obj}

\State Set constraints \eqref{eq:lim_1}-\eqref{eq:trans_stab_const_lim_9} 

\State Solve to get UC status, $u_i$

\end{algorithmic}
\end{algorithm}

\subsection{Corrective Optimal Power flow (OPF)}

In real-time, the rescheduling must be done 
according to the updated forecasts and contingency risk.
By quantifying the risk of a wildfire affecting the region, we can use it to relax constraints \eqref{eq:lim_6}-\eqref{eq:trans_stab_const_lim_9} which may be too limiting otherwise, 
to solve a modified OPF problem (called the CSCOPF problem) and get the desired redispatch. The modified objective is now expressed as:
\begin{equation}
\label{eq:CSCOPF_obj}
    \begin{aligned}
    \min_{\Delta p_{i}, \Delta l_j} \quad &  \sum_{\forall i \in G} (c_i \Delta p_{i}^2 + d_i \Delta p_{i}) + \sum_{\forall j \in L} (m_j \Delta l_j)\\
     \quad & + \lambda_c ( \sum_{\forall i \in G} \sum_{\forall u \in K_{\mathrm{crit}}} \mathrm{PTDF}_{u,i} \Delta p_{i} \\
         \quad & - \sum_{\forall j \in L} \sum_{\forall u \in K_{\mathrm{crit}}} \mathrm{PTDF}_{u,j} \Delta l_j + \Delta P_{K_{\mathrm{crit}}}) \\ 
    \quad & + \lambda_t (\sum_{\forall i \in \mathrm{CM}} \Delta p_{i}  + \Delta P_{tr}) +  \lambda_N (\sum_{\forall e \in \xi} f_e^2 )
    \end{aligned}
\end{equation}
\nomenclature{\( \lambda_N \)}{Risk metric for power flowing through vulnerable lines}
\nomenclature{\( \lambda_t \)}{Risk metric for transient stability correction}
\nomenclature{\( \lambda_c \)}{Risk metric for cut-set desaturation}
\nomenclature{\( \lambda_b \)}{Risk metric for post-contingency branch overloads}
\nomenclature{\( \lambda \)}{Generalised wildfire risk in target area}

In \eqref {eq:CSCOPF_obj}, the constraint penalties $\lambda_c$ and $\lambda_t$ are representative of the wildfire risk in the target area, and can be varied appropriately to enforce compliance of the cut-set and transient stability constraints.
The above-mentioned objective is now subjected to the following constraints, which are similar to those specified for the CSCUC with the additional inclusion of the constraint penalties:
\begin{equation}
\label{eq:CSCOPF_C1}
    p_{i}^{\mathrm{min}} - p_{i}^0 \geq \Delta p_{i} \geq p_{i}^{\mathrm{max}} - p_{i}^0 \;  \forall i \in G
\end{equation}
\begin{equation}
    \label{eq:CSCOPF_C2}
    l_j^{\mathrm{min}} - l_j^0 \geq \Delta l_j \geq l_j^{\mathrm{max}} - l_j^0 \quad \forall j \in L
\end{equation}
\begin{equation}
    \label{eq:CSCOPF_C3}
\begin{aligned}
     \quad & f_e^{\mathrm{min}} - f_e^0 \leq \sum_{\forall i \in G} \mathrm{PTDF}^r_{e,i} \Delta p_{i} - \sum_{\forall j \in L} \mathrm{PTDF}^r_{e,j} \Delta l_j \\
      \quad & \leq f_e^{\mathrm{max}} - f_e^0   \quad \forall e \in B_r\\
\end{aligned}
\end{equation}
\begin{equation}
    \label{eq:CSCOPF_C4}
    \sum_{\forall i \in G} \Delta p_{i} = \sum_{\forall j \in L} \Delta l_j
\end{equation}
\begin{equation}
\label{eq:CSCOPF_C5}
\begin{aligned}
    \quad & \sum_{\forall i \in G} (\mathrm{PTDF}^r_{e,i} + \mathrm{LODF}_{e,k} \mathrm{PTDF}^r_{k,i}) \Delta p_{i}\\
     - \quad & \sum_{\forall j \in L} (\mathrm{PTDF}^r_{e,j} + \mathrm{LODF}_{e,k} \mathrm{PTDF}^r_{k,j}) \Delta l_j\\
    \leq \quad & \lambda_b(f_e^{\mathrm{max}} - f_e^0 + (\mathrm{LODF}_{e,k} f_k^0)) \quad \forall e,k \in B_r, \xi
\end{aligned}
\end{equation}
\begin{equation}
\label{eq:CSCOPF_C6}
\begin{aligned}
    \quad & \sum_{\forall i \in G} (\mathrm{PTDF}^r_{e,i} + \mathrm{LODF}_{e,k} \mathrm{PTDF}^r_{k,i}) \Delta p_{i} \\
     - \quad & \sum_{\forall j \in L} (\mathrm{PTDF}^r_{e,j} + \mathrm{LODF}_{e,k} \mathrm{PTDF}^r_{k,j}) \Delta l_j\\
    \geq \quad & \lambda_b (f_e^{\mathrm{min}} - f_e^0 + (\mathrm{LODF}_{e,k} f_k^0)) \quad \forall e,k \in B_r, \xi
\end{aligned}
\end{equation}

The CSCOPF problem formulated above utilizes the status of the generators calculated in the CSCUC as inputs. This ensures that the resulting redispatch is fast as well as economical. 
The values of the constraint penalties, denoted by $\Lambda = [\lambda_b \quad \lambda_c \quad \lambda_t \quad \lambda_N]$ which constitute the real-time wildfire risk, may be obtained from predictive models using external factors such as weather data \cite{malik2021data}, or from grid-edge devices monitoring for faults or smoke \cite{allison2016airborne}; it may also be set manually from operator experience. 
Furthermore, its value may be altered in real-time to shift the system to a more alert state or less alert state depending upon the probability of wildfire in the region-under-study. 

Algorithm \ref{alg:CSCOPF} describes the implementation of CSCOPF. It takes inputs from various sources, including the trained TSCP $\Upsilon$, and the allowed wildfire risk, $\Lambda$. The warm start solution $p_i^0$ may be obtained from the real-time power injections (which can be estimated \cite{sahoo2023data}) and the load forecasts. Once the risk metrics are set, CSCOPF is run, and the resulting solution is checked for any additional violations using FT and a single TDS. 
As one approaches the most vulnerable hours of the day,
the value of $\Lambda$
should be increased to 
better prepare the system for facing the wildfire contingency.

\begin{algorithm}
\caption{CSCOPF Implementation}\label{alg:CSCOPF}
\textbf{Input:} Units committed $u_i$, Real-time power injections, Real-time wildfire risk $\Lambda$, Transient stability analyser $\Upsilon$ \\
\textbf{Output:} optimal redispatch $\Delta p_i$, and load shed $\Delta l_j$\\
\textbf{Real-time:}
\begin{algorithmic}[1]
\State Obtain or set risk $\lambda$
\State Define empty list of constraints $\Phi$
\State Obtain real-time 
power injection information
    \State Generate cut-set constraint using \eqref{eq:lim_8}
    \State Generate transient stability constraint using $\Upsilon$; see \eqref{eq:trans_stab_const_lim_9}
    \State Update $\Phi$ with constraints \eqref{eq:lim_8} and \eqref{eq:trans_stab_const_lim_9}
\While{additional violations detected}

    \State CSCOPF: Define objective \eqref{eq:CSCOPF_obj} using $\Phi$, apply
    \Statex $\quad \;$ constraints \eqref{eq:CSCOPF_C1}-\eqref{eq:CSCOPF_C6} \label{Step}
    \State Solve CSCOPF to get $\Delta p$ and $\Delta l$
    \State Update dispatch and run power flow
    \State Run FT and single TDS on updated dispatch
    \If {violations detected}
        \State Update $\Phi$ and Go to Step \ref{Step}
    \EndIf
\EndWhile
    
\end{algorithmic}
\end{algorithm}

\section{Results}\label{Section4}

\begin{figure}[b]
\includegraphics[width=0.485\textwidth]{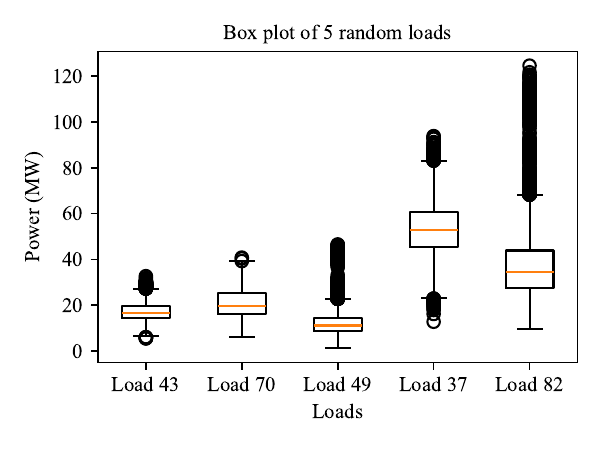}
\caption{{Load variation data for five loads in the 118-bus system}}
\label{fig:load_var}
\end{figure}

The proposed formulation is tested on the publicly available IEEE 118-bus system \cite{IIT118bus2003}. The system consists of 118 buses, 54 generators, and 99 loads. This system is known to be very robust and stable against line outages, so the effects of simulated wildfires in this system emphasizes the true effects of the damage an unmitigated approach can have.
For generating diverse operating conditions, equivalent loads 
from the publicly available 2000-bus synthetic Texas system \cite{birchfield2017grid} are first found, and then their variations were captured using kernel density estimation (KDE). 
The variations in some of the loads are shown in Fig. \ref{fig:load_var}. Random sampling was then performed from the KDEs to generate $\approx28000$ samples of real power load $l_i$.
These loading conditions were then used to perform contingency analysis using TDS.
The TDSs were performed in $\mathrm{PSSE^\circledR}$, while the two-stage optimization was modeled and solved using $\mathrm{Gurobi}$ and $\mathrm{Pandapower}$ in $\mathrm{Python~3.11}$.



A common theme across Vignettes 1 and 2 is the possibility of a wildfire impacting a region of the power system in the near-future (e.g., the next day).
Accordingly, a prospective transmission corridor comprising two lines (namely, $23-25$ and $26-30$) of the 118-bus system was identified, and 
it was assumed that on the following day, a set of five arc-faults would occur consecutively over a period of three seconds at the end of which each line would suffer a permanent outage.
The corrective action analysis used to generate the power shift required in \eqref{eq:trans_stab_const} for all the loading conditions were then calculated, and $\Upsilon$ was trained, with the train to test split in the ratio $80:20$. 
For obtaining the results shown in the next four sub-sections, the constraint penalties/risk metrics were made to obey the following relation: $\lambda_b =  \lambda_c = \lambda_t = \lambda_N * 0.01 = \lambda$; this condition was relaxed in Section \ref{sec:Sensitivity_analysis}.
Note that $\lambda_N$ includes a normalising factor as the associated term is of the unit $MW^2$.
All computational analyses done in this paper were performed using a computer with an Intel Core (TM) i7-11800H CPU @2.3GHz with 16GB of RAM and an RTX 3070Ti GPU.
 


\subsection{Contingency Alleviation Action}
\label{Section4A}
The cut-set and transient stability constraints generated for this extreme event scenario is detailed in \autoref{tab:cont_solns}. 
The constraints also define the instability alleviation actions, namely, the dispatch of the critical generators $(25, 26)$ must be reduced by at least 118 MW, and the aggregate power flow across the lines $(26-30, 25-27)$ must be reduced by at least 187 MW.
Note that the required rescheduling is considerably large, implying
that the alleviation will 
be subject to ramp-rate constraints. Furthermore, such a large change does lead to additional security violations for the test system,
which are captured by the FT algorithm. For these (as well as economic) reasons, it is essential that 
a coordinated strategy involving both day-ahead and real-time control are employed instead of entrusting everything to real-time corrections.

\begin{table}[ht]
\centering
\caption{Contingency Impacts}
\label{tab:cont_solns}
\begin{tabular}{|l|c|}
\hline
\textbf{System}                                                                                & \textbf{118-bus system}  \\ \hline

Lines tripped & (23-25, 26-30) \\ \hline

$\mathrm{N-1}$ branch overloads identified & (8-5) \\ \hline
{Saturated cut-sets }                                                                   &     (26-30, 25-27)                                              \\ \hline
{ $\Delta P_{K_{\mathrm{crit}}}$ }                                                               &       187.086 MW                                         \\ \hline
{\% of capacity }                                                                        &          42.51\%                                       \\ \hline
{\begin{tabular}[c]{@{}c@{}}
Critical Machines ($\mathrm{CM}$) for transient stability\end{tabular}} &        25, 26                                         \\ \hline
{$\Delta P_{tr}$}                                                                                  &         118 MW                                        \\ \hline
Total capacity of tripped generation & 534 MW \\ \hline
{\% of capacity}                                                                        &          22.1\%                                        \\ \hline
\end{tabular}
\end{table}

\begin{figure}
\centering
\begin{subfigure}[b]{.485\textwidth}
  \centering
  \includegraphics[width=\textwidth]{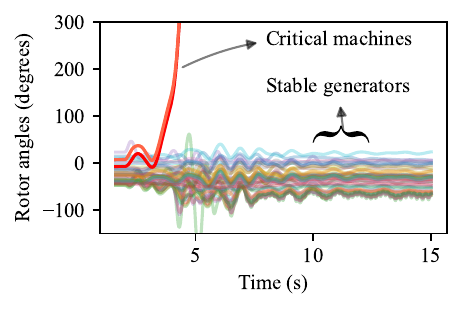}
  \caption{{TDS without control}}
  \label{fig:sub1}
\end{subfigure}%
\\
\begin{subfigure}[b]{.485\textwidth}
  \centering
  \includegraphics[width=\textwidth]{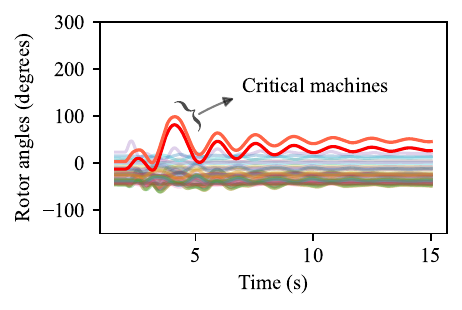}
  \caption{{TDS with control}}
  \label{fig:sub2}
\end{subfigure}
\caption{{Rotor angle stability for the 118-bus system}}
\label{fig:118_rotorangles}
\end{figure}

For the contingency-under-investigation, the proposed contingency analysis tool was able to quantify the impacts (of the contingency) as well as determine the actions that must be taken to alleviate them. 
As an example, Fig. \ref{fig:118_rotorangles} illustrates the transient stability (rotor angles) of two TDSs for this contingency
before and after implementing the recommended alleviation actions. The unstable generators $(25, 26)$ are able to ride-through the contingencies and do not lose synchronism after the corrections are applied (compare the red and orange-colored curves in this figure).

The results for the TSCP algorithm,
$\Upsilon$, is given in \autoref{tab:shallow_models}. 
In this study, $\Upsilon$ trained using a linear regression model is compared with other data-driven models including ridge regression, support vector regression (SVR) with a radial basis function (RBF) kernel, XGBoost, random forest, elastic net, $k$-nearest neighbor (KNN), decision tree, and least absolute shrinkage and selection operator (LASSO) regression. 
The metrics used to gauge the performance of these models include the root mean squared error (RMSE),  $R^2$ score, mean bias deviation ($\mathrm{MBD}$), and $R^2$ robustness.
Note that $\mathrm{MBD}$ is a representation of the average bias in the model predictions, and is mathematically defined as:
\begin{equation}
    \mathrm{MBD} =  \frac{\sum_{n} (y - \hat{y})}{n}
\end{equation}

Ideally, $\mathrm{MBD}$ should be 0, to prevent bias in estimation. However, for the identified problem, underestimation of TSCF (positive $\mathrm{MBD}$) is much worse than overestimation, as the latter leads to uneconomical operation, while the former leads to transient instabilities and the eventual tripping of generators. 
Therefore, data-driven models that give a \textit{small negative} value for $\mathrm{MBD}$ are better.
Finally, since load forecasts are naturally subject to errors, it is important to also consider the robustness drop in $R^2$, which is calculated by introducing noise into the forecasts and observing the subsequent change in the $R^2$ score. Naturally, a smaller value of $R^2$ robustness is preferred. 

\begin{table}[ht]
\centering
\caption{TSCP Results for Different Data-driven Models}
\label{tab:shallow_models}
\begin{tabular}{|l|c|c|c|c|c|c|c|c|c|}
\hline
\textbf{Model} &

\begin{tabular}{c}
    \textbf{RMSE}  \\
    (MW)
\end{tabular}

& \textbf{$R^2$}  & \textbf{MBD}   & \begin{tabular}{c}
    $R^2$  \\
    \textbf{Robust-} \\
    \textbf{ness}
\end{tabular}\\
\hline
Linear Regression & 0.31 & 0.98 &\(-0.57\mathrm{e}{-4}\)  & 0.0024 \\
\hline
Ridge Regression & 0.31 & 0.98  &\(-0.57\mathrm{e}{-4}\)   & 0.0027 \\
\hline
SVR & 0.42 & 0.97  & 0.02  & 0.0028 \\
 \hline
 XGBoost & 0.94 & 0.85  & -0.004  & 0.0035 \\
\hline
Random Forest & 1.51 & 0.61  & 0.005   & 0.0031 \\
\hline
Elastic Net & 2.13 & 0.22 & \(-0.93\mathrm{e}{-4}\)   & -0.0003 \\
\hline
KNN & 2.21 & 0.16  & 0.58  & -0.0004 \\
\hline
Decision Tree & 2.33 & 0.06  & 0.002 & -0.0051 \\
\hline
LASSO Regression & 2.34 & 0.06  & \(-0.28\mathrm{e}{-4}\)
 & 0.0001 \\
\hline
\end{tabular}
\end{table}

It is clear from the table that linear regression has a low error with a high degree of confidence, even in the presence of noise in the forecasts.
This result confirms our earlier stipulation
that 
a  quasi-linear relationship exists between the pre-contingency loading conditions and the transfer margin (see Section \ref{ML-TSA}). 
Moreover, the shallow model (linear regression) combined with the interpretable nature of the constraints ensures transparency, i.e, should prevailing conditions shift, a power system operator possesses the discretion to intervene at any stage of the implementation, allowing the application of custom solutions derived from their experience.

\subsection{Day-ahead Unit Commitment }

\textcolor{black}{The contingency analysis results shown in \autoref{tab:cont_solns} are now used for optimal decision-making.}
In anticipation of a wildfire impacting the identified transmission corridor of the 118-bus system in the following day, the
proposed modified UC is performed
to determine and dispatch additional generators.
The CSCUC is solved assuming the contingency has already manifested, and in the post-contingency state, the committed generators are to be dispatched.
The results obtained are shown in \autoref{tab:CSCUC_Results}.
The CSCUC recommends bringing an additional generator online at bus 6. Note that this generator may not dispatch 
during the times of the day when the risk of fire is low. However, it is cheaper and faster (as shown later in 
Section \ref{Comparison})
to bring the system to a ready state with this generator in-service when the risk of wildfire is high (Vignette 1) or the fire enters the 
region-under-study (Vignette 2).


\begin{table}[ht]
\centering
\caption{CSCUC Results}
\label{tab:CSCUC_Results}
\begin{tabular}{|l|c|}
\hline
\textbf{Parameter}              & \textbf{Value}        \\ \hline
Contingency List                & {[}(23-25), (26-30){]} \\ \hline
Number of additional generator(s) committed & 1                     \\ \hline
Generator(s) committed            & {Generator at bus [}6{]}               \\ \hline
Total objective (\$/hr)            & 665.31                \\ \hline
Time to solve (s)               & 0.596                 \\ \hline
\end{tabular}
\end{table}

\subsection{Real-time Optimal Redispatch}
The optimal reschedule primarily depends on the {real-time} status of the wildfire: whether it actually occurs in case of Vignette 1 or enters the region-under-study in case of Vignette 2, or it has not impacted the region-under-study yet but the risks continue to be high. It also depends on how alert the power system operator wants the system to be (i.e., the level of allowed risk specified by $\lambda$).
In the case when a fire is in the region-under-study, the CSCOPF solution is obtained by solving \eqref{eq:new_obj} with the values of $u_i$ set based on the day-ahead CSCUC. 
In the case when a fire has not started but the risks are high, the risk-based formulation given in \eqref{eq:CSCOPF_obj} allows the system to increase its readiness to the contingency gradually (e.g., by increasing $\lambda$).

\begin{figure}[ht]
	\includegraphics[width=0.485\textwidth]{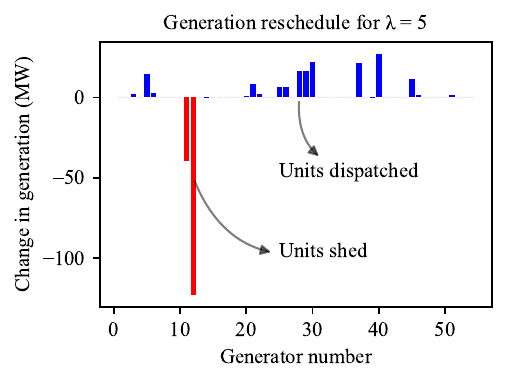}
	\caption{{CSCOPF result: Generator reschedule. Blue generators take on additional generation, while red generators lower their dispatch}}
	\label{fig:CSCOPF_result_1}
\end{figure}

In the following results, we consider the latter case in which the fire has not started in the region-under-study of the 118-bus system but the risks remain high.
A reschedule of the generators based on $\lambda=5$ is shown in Fig.  \ref{fig:CSCOPF_result_1}. In the figure, the blue bars indicate generators that take on additional generation, while the red bars indicate generators that lower their active generation. The major reduction in generation is seen in generators which are directly affected by the contingency-under-consideration, and  CSCOPF reschedules them economically while maintaining the security of the rest of the system.



\begin{table*}[ht]
\centering
\caption{{Comparative Analysis of the Proposed Model with State-of-the-art}}

\begin{tabular}{|l|c|c|c|c|c|c|c|c|}
\hline

\multirow{2}{*}{\textbf{Metric}} & \multicolumn{3}{c|}{\textbf{Two-stage CSCUC/CSCOPF}} & \multicolumn{1}{c|}{\multirow{2}{*}{\textbf{RT-SCED}}} & \multicolumn{1}{c|}{\multirow{2}{*}{\textbf{TSCOPF} \cite{10138159}}} & \multicolumn{1}{c|}{\multirow{2}{*}{\textbf{CSCOPF} \cite{sahoo2023cut}}} & \multicolumn{1}{c|}{\multirow{2}{*}{\textbf{CSCUC/OPF}}} \\ \cline{2-4}
 & \textbf{$\lambda$ = 0.1} & \textbf{$\lambda$ = 5} & \textbf{$\lambda$ = 7} & & & & \\ \hline

\textbf{Contingency status} & \multicolumn{3}{c|}{\textit{Pre-contingency}} & \multicolumn{4}{c|}{\textit{Post-contingency}} \\ \hline
Power shed in vulnerable lines (\%) & 12.52 & 14.95 & 16.32 & 100 & 100 & 100 & 100 \\ \hline
CM Generation Shed (MW) & 17.73 & 106.88 & 140.82 & 20.4 & 118 & 269.79 & 310.88 \\ \hline
Cut-set desaturation (MW) & 21.43 & 61.93 & 76.41 & -131.80 & 57.52 & 187.08 & 187.08 \\ \hline
Total load shed (MW) & 0 & 0 & 0 & 0 & 0 & 0 & 0 \\ \hline
\textbf{Transient stable?} & \textit{No} & \textit{No} & \textit{Yes} & \textit{No} & \textit{Yes} & \textit{Yes} & \textit{Yes} \\ \hline
\textbf{Cut-set secure?} & \textit{No} & \textit{No} & \textit{No} & \textit{No} & \textit{No} & \textit{Yes} & \textit{Yes} \\ \hline
Additional generators committed & 1 & 1 & 1 & 0 & 0 & 0 & 1 \\ \hline
Number of dispatch cycles to clear constraints & 3 & 2 & 1 & 0 & 0 & 0 & 0 \\ \hline
Additional real-time cost (\$/hr) & 447.81 & 623.83 & 760.04 & 7,526.11 & 7,288.42 & 9,254.09 & 8,353.83 \\ \hline
Time to solve (s) & 0.662 & 0.662 & 0.662 & 0.066 & 0.256 & 0.066 & 0.662 \\ \hline
\end{tabular}

\label{tab:Full_result}
\end{table*}

\subsection{Comparison with State-of-the-Art}
\label{Comparison}

A comparison of the proposed model with the conventional real-time security constrained economic dispatch (RT-SCED), a traditional transient stability constrained optimal power flow (TSCOPF) model \cite{10138159}, {and our previous CSCOPF formulation in \cite{sahoo2023cut}} is performed for the identified contingency in the 118-bus system, and the results are shown in \autoref{tab:Full_result}. 
Note that the RT-SCED model has no cut-set or stability constraints, while the TSCOPF has no cut-set constraints.
Similarly, the CSCOPF developed in \cite{sahoo2023cut} only operates in the post-contingency stage in which the fire is already in the region-under-study.


In the table, both the pre-contingency and the post-contingency 
scenarios under \textit{vignettes 1 and 2} are considered.
In the day-ahead stage, the proposed coordinated CSCUC/OPF formulation is implemented to dispatch an additional generator (at bus 6) which is used in real-time to reduce system vulnerability by controlling the risk parameters.
While the ideal value of $\lambda$ is subject to operator experience, one objective of tuning it would be to minimize the time required to remove system vulnerabilities, ideally bringing it to within a single dispatch cycle. 
For example, operating at $\lambda = 0.1$ would be cheaper than operating at $\lambda = 7$. However, upon contingency manifestation or during periods of high risk when the utility decides to preemptively shut the region down, it would take at least three times longer to bring the system into a secure and stable state using $\lambda = 0.1$ vs. $\lambda = 7$; the difference arising due to the ramp rates of the various generators.
In the post-contingency stage, while TSCOPF was able to alleviate the stability constraint, it did not alleviate the cut-set insecurities. 
Meanwhile, RT-SCED increased the cut-set saturation, making the region even more vulnerable to additional asset loss.
Conversely, the CSCOPF developed in \cite{sahoo2023cut} and the proposed CSCOPF/OPF are able to alleviate both the vulnerabilities, resulting in a static-and-dynamically secure solution without resorting to any load shed. 
However, the proposed model results in a lower additional cost as elaborated below.

{The additional operational cost (henceforth referred to as simply \textit{cost}) under all conditions relative to base dispatch given the same load forecasts are given in the second-to-last row of the table.}  For context, a system with no control/an insecure solution would have put about 534 MW of generation at risk of tripping which would have led to a loss in revenue of at least \$12,651.44/hr, calculated on the generation side.
{In the pre-contingency stage, the real-time cost is significantly lower than the cost incurred for all post-contingency scenarios, which shows the advantages of using the proposed approach when the risks are relatively lower, or if the wildfire has not started yet. 
Note that for $\lambda=7$, the system can be operated with the transient stability constraint cleared (the CMs will not lose synchronism even under the worst-case scenario) while the cost is about one-tenth of the (economically) best case post-contingency operation (\$760.04 compared to \$7526.11 obtained for RT-SCED). 
Now, performing an economic comparison of the proposed CSCUC/OPF with the CSCOPF developed in \cite{sahoo2023cut} for the post-contingency stage, it is clear that the cost of the former is \$900.26/hr lower than that of the latter (\$8,353.93/hr for the proposed model and \$9,254.09/hr for the model developed in \cite{sahoo2023cut}).
This difference in cost shows the importance of a two-stage preventive-corrective coordinated approach.} 
Essentially, while the proposed solution leads to increased start-up costs due to the dispatch of an additional generator (at bus 6), the costs are recovered 
through lower price of real-time operation.

{It is also important to consider the computational burden of the optimization to ensure feasibility during real-time operation; this information is  provided in the last row of \autoref{tab:Full_result}.
The solution time for the proposed model is presented as the total time to solve both the UC and OPF models, albeit they are run at different times. The UC's model complexity is bounded by only the set of generators that are \textit{not} dispatched, and the security and stability criteria are formulated as linear constraints of the decision variables, which aims to minimize solution time. For the 118-bus system, the solve time of the CSCUC model was 0.596s, and the CSCOPF model was 0.066s, bringing the total to 0.662s.}
The solve time of CSCOPF is similar to RT-SCED and one-fourth of TSCOPF, implying that the additional constraints do not significantly increase the computational burden of the optimization, with the data-driven transient stability analyzer improving the speed of online operation, { as compared to the TDS required for conventional TSCOPF.}
Lastly, although load-shed was not required for the contingency-under-study, it is important to incorporate it in the problem  formulation {to increase operational flexibility.}

\vspace{-1em}

\subsection{
Sensitivity Analysis}\label{sec:Sensitivity_analysis}

This sub-section explores the flexibility in implementation of the proposed model under 
the identified problem scope for the region-under-study of the 118-bus system. The primary purpose is to understand the full scale of its 
applicability and simultaneously, its limitations.
To do this, we evaluate the performance of the proposed two-stage optimization formulation by varying $\lambda$, i.e, for varying wildfire risk situations.

For a small $\lambda$, economical dispatch is preferred, while the vulnerability of the system to contingencies is high, as it would take a longer amount of time for the critical generators to ramp down to stable levels of generation. Note that one dispatch cycle is considered to be 15 minutes here. Meanwhile by increasing $\lambda$, the model prefers clearing these constraints, and hence the risk of the system being affected by a wildfire reduces significantly, albeit at a higher operational cost indicated by the higher (positive) value of the CSCOPF objective.



\begin{figure}[b]       \includegraphics[width=0.485\textwidth]{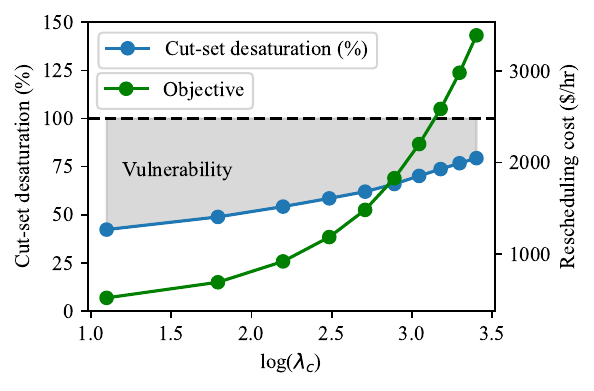}

	\caption{{Sensitivity analysis of the cut-set risk $\lambda_c$}}
	\label{fig:cutset_sensitivity_lamda}
\end{figure}
\begin{figure}
	\includegraphics[width=0.485\textwidth]{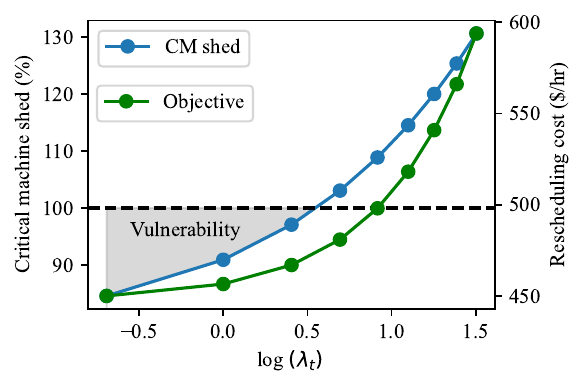}
	\caption{{Sensitivity analysis of the transient stability risk $\lambda_t$}}
	\label{fig:TS_sensitivity_lamda}
\end{figure}

A sensitivity analysis for the most important individual risks is explored next. The sensitivities for the risk metric $\lambda_c$ and $\lambda_t$ are shown in Figs. \ref{fig:cutset_sensitivity_lamda} and \ref{fig:TS_sensitivity_lamda}, respectively. In both the figures, the identified vulnerable regions are the regions where the system may be operated at if the wildfire has not manifested yet, for economic reasons, as reducing vulnerability is associated with an almost equivalent increase in cost. 
\textcolor{black}{These analyses have different effects based on the problem scope. For situations of active wildfire risks in which the contingency has not manifested yet, but the risk is very much present, the system may be run in the vulnerable regions of Figs. \ref{fig:cutset_sensitivity_lamda} and \ref{fig:TS_sensitivity_lamda}. For example, 
a recommended solution is to set $\mathrm{log}(\lambda_c) = 2.6$ and $\mathrm{log}(\lambda_t) = 0.4$, as it results in a reasonable trade-off between operational cost and system vulnerability. On the contrary, if there is a fire in the region-under-study, the system must be operated beyond the vulnerability regions for both the risk metrics since the affected lines would have to be tripped}.
Lastly, note that in all of these results,
the total load was served, which demonstrates that it is possible to increase supply reliability without increasing system vulnerability. 

\begin{figure}[ht]
	\centering
	\includegraphics[width=0.485\textwidth]{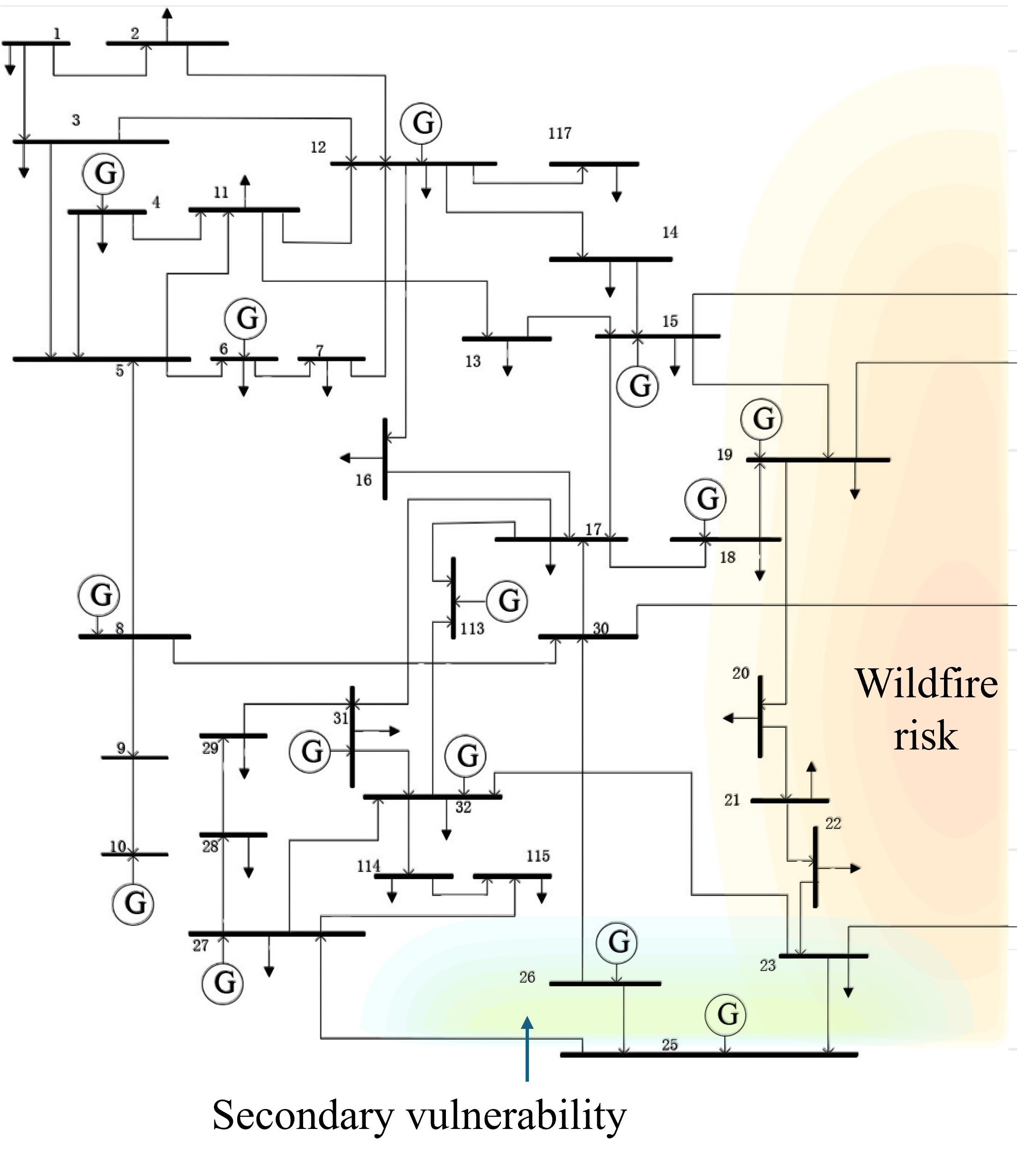}
	\caption{{Case-study of an islanding contingency in IEEE 118-bus system. ''Wildfire risk" denotes the region of the system where the contingency occurs, while the ''secondary vulnerability" denotes a new region of the islanded system that becomes vulnerable after the island has formed}}
	\label{fig:casestudy}
\end{figure}

{
\subsection{Case Study: Severe Islanding Contingency}
The primary objective of the proposed CSCUC/OPF formulation is to mitigate the worst-case impact and reduce system vulnerabilities caused by a system disruption, whether it is planned (preemptive shutoffs) or unplanned (after a wildfire has already manifested). However, if the risks are too high, utilities resort to PSPS to avoid the grid itself from starting a fire. 
In extreme cases, this may lead to controlled islanding of the network to prevent widespread disruptions.
This case study models such an islanding case in the 118 bus system, and analyzes the effect of the proposed formulation in both the pre-contingency and the post-contingency stages.
The location of the island and the affected lines are shown in Fig. \ref{fig:casestudy}.
The affected lines span across a distinct corridor which carves out an
island that is $\approx$ 30\% of the 
network. The intertie consists of four lines (15-33, 19-34, 30-38, and 23-24), with a total transfer capacity of  1.4 GW.
Islanding of this region may happen either: if the forecasted wildfire risk is very high, and the utilities intend to conduct PSPS \textit{(Vignette 1)}, or if a wildfire breaks out in the vulnerable region \textit{(Vignette 2)}.
We analyze this case-study from two perspectives as elaborated below.}

{
\subsubsection{Ensuring post-contingency stability}
In this sub-section, the contingency analysis tool is used to assess the resilience of the system due to the eventual creation of the island. While there were no cut-set or transient stability violations for the outage/de-energization of the lines, the tool identified violations in a \textit{secondary} region of the created island, the details of which are given in the contingency analysis portion of \autoref{tab:casestudy}.
The violations include a vulnerable cut-set across three lines (23-25, 25-27, and 26-30) and two vulnerable generators (at buses 25 and 26); the corresponding stability margins/entries are also provided in the table. Note that upon the creation of the island, all the power flowing through the de-energized lines is now redirected through the vulnerable
secondary
cut-set, which increases the static insecurity of the island. If any assets are now affected by a fire spreading into the secondary region \textit{(Vignette 2)},
it may lead to an outage of the entire vulnerable
region and the corresponding
generators. 
This shows that the proposed model can be used to provide insights into the stability of the grid
in
the event of an island formation or reduced network operation.
Furthermore, by alleviating this vulnerability using the proposed model (described in the second half of \autoref{tab:casestudy} and elaborated in the next sub-section),
reliable power dispatch within the entire grid can be ensured.
}

\begin{table}[ht]
\caption{{Case study: Islanding results}}
\begin{tabular}{|lc|}
\hline
\multicolumn{1}{|l|}{\textbf{Property}}                                                                       & \textbf{Value}                     \\ \hline
\multicolumn{2}{|c|}{\textit{Contingency Analysis}}                                                                                                \\ \hline
\multicolumn{1}{|l|}{Lines shut down/affected}                                                                & (15,33), (19,34), (30,38), (23,24) \\ \hline
\multicolumn{1}{|l|}{\begin{tabular}[c]{@{}l@{}}Power flow across\\  affected lines (MW)\end{tabular}}        & 77.48                              \\ \hline
\multicolumn{1}{|l|}{Size of island formed}                                                                   & 35 buses                           \\ \hline
\multicolumn{1}{|l|}{Vulnerable cut-set detected}                                                             & (23, 25), (25, 27), (26, 30)       \\ \hline
\multicolumn{1}{|l|}{Cut-set transfer margin (MW)}                                                            & 77.48                              \\ \hline
\multicolumn{1}{|l|}{Vulnerable generators}                                                                   & 25, 26                             \\ \hline
\multicolumn{2}{|c|}{\textit{CSCUC/OPF ($\lambda$=7)}}                                                                                                   \\ \hline
\multicolumn{1}{|l|}{Additional generator dispatched}                                                         & Generator at bus {[}40{]}                           \\ \hline
\multicolumn{1}{|l|}{Load shed (MW)}                                                          & 0                             \\ \hline
\multicolumn{1}{|l|}{Affected lines power shed (MW)}                                                          & 22.21                              \\ \hline
\multicolumn{1}{|l|}{\begin{tabular}[c]{@{}l@{}}Vulnerable cut-set \\ desaturation (MW)\end{tabular}}         & 33.9                               \\ \hline
\multicolumn{1}{|l|}{\begin{tabular}[c]{@{}l@{}}Additional real-time\\ operational cost (\$/hr)\end{tabular}} & 121. 53                            \\ \hline
\multicolumn{1}{|l|}{Dispatch cycles to shut-down}                                                            & 1                                  \\ \hline
\end{tabular}
\label{tab:casestudy}
\end{table}

{
\subsubsection{Facilitating preventive PSPS operation}
Although the post-contingency stability of the island is assessed by considering the static and/or dynamic insecurities created by asset loss after the island is created, it is better to alleviate these vulnerabilities \textit{before} the island is formed; this aspect is analyzed in this sub-section.
If the utility preemptively conducts PSPS de-energizing the four lines in the wildfire risk area \textit{(Vignette 1)}, the proposed model can be used to alleviate post-contingency vulnerabilities by reducing the power flowing through the high risk region (while causing minimal addition to operational cost), so that the region can be quickly shut down when the risk is at its peak.
The results are shown in the second half of  \autoref{tab:casestudy}. Implementation of the proposed coordinated CSCUC/OPF dispatched an additional generator at  bus 40 in the day-ahead stage. In real-time, before the system risks increase dramatically (thus forcing PSPS), the system is run in an intermediary stage, where the power flowing through the vulnerable area (secondary cut-set in this case) is reduced just enough to enable the utility to disconnect the four lines within one dispatch cycle, while ensuring that the security and stability of the island is maintained, while also not causing any load shed.
For comparison, doing PSPS alone
would have required over two dispatch cycles, while increasing operational cost post shutdown as well.
In summary, from this case study, it is observed that the proposed model can be used to increase system resilience during wildfire events, while also
facilitating the PSPS implementation process.}   

\section{Conclusion}\label{Section5}

In this paper, a holistic preventive-corrective action scheme addressing both static and dynamic security criteria was introduced that facilitates stability and security-constrained unit commitment (UC)/optimal power flow (OPF) solutions for wildfire-related contingencies, and periods of active wildfire risk. 
\textcolor{black}{A fast and scalable algorithm for cut-set detection and correction, called Feasibility Test (FT), is used to analyze extreme event scenarios from a static security perspective, and prevent cascading line outages.
This algorithm is combined with a}
 data-driven linear transient stability  constraint prediction (TSCP) model that is able to accurately and reliably predict the appropriate correction factor for mitigating transient instabilities (caused by generator outages) under various loading conditions.
 
 \textcolor{black}{
A two-stage optimization model is then proposed that utilizes the contingency analyses and enables optimal decision-making in the day-ahead and real-time stages. In the day-ahead stage, the cut-set and stability constrained unit commitment (CSCUC) performs a modified
UC considering a post-contingency scenario to determine additional generators to be brought online. In real-time, the cut-set and stability constrained optimal power flow (CSCOPF) is able to determine multiple redispatch solutions based on the varying real-time risk of wildfires.
}

Implementation of the proposed model on the IEEE 118-bus system shows that the FT algorithm and the TSCP model are able to accurately and reliably predict the static and dynamic stability constraints in the face of load uncertainties. 
The numerical results show that the proposed preventive-corrective coordinated optimization is able to detect and alleviate all cascading outage risks for both lines and generators, while bearing minimal additional operational cost. A comparison with similar models shows that the proposed model is able to achieve security and stability while being fast and economical. 
With suitable modifications, the proposed formulation can be applied to other extreme event scenarios as well.
{Future work will pivot towards the synergistic integration of renewable resources, from both economic and physical perspectives (such as ride-through constraints), into the proposed CSCUC/CSCOPF problem formulation.}

\bibliography{bibtex.bib}

\end{document}